\documentclass[12pt,a4paper]{article}
\usepackage[utf8]{inputenc}
\usepackage{jheppub}
\usepackage{color}
\usepackage[table]{xcolor}
\usepackage{amsmath,amssymb}
\usepackage{graphicx,graphics}
\usepackage{slashed}
\usepackage{url}
\usepackage{enumitem}
\usepackage{multirow}
\usepackage{array}
\usepackage{tabulary}

\begin{document}
\preprint{HRI-RECAPP-2021-011}
%%%%%%%%%%%%%%%%%%%%%%%%%  Title  %%%%%%%%%%%%%%%%%%%%%%%%%%
\title{Jet Substructure and Multivariate Analysis Aid in
Polarization Study of Boosted, Hadronic $W$ Fatjet at the
LHC}

%%%%%%%%%%%%%%%%%%%%%%%%%  Authors  %%%%%%%%%%%%%%%%%%%%%%%%
\author[a]{Atri Dey,}
\emailAdd{atridey@hri.res.in}
\affiliation[a]{Regional Centre for Accelerator-based Particle Physics, %\\ 
Harish-Chandra Research Institute, HBNI,
Chhatnag Road, Jhunsi, Prayagraj (Allahabad) 211\,019, India}

\author[a]{and Tousik Samui}
\emailAdd{tousiksamui@hri.res.in}

%%%%%%%%%%%%%%%%%%%%%%%%%  Abstract  %%%%%%%%%%%%%%%%%%%%%%%
\abstract{Study of polarization of heavy particles is an
important branch of research in today's collider studies.
The massive $W$ boson has two types of polarization states,
which usually are studied via the angular distribution of
its decay products. We have studied polarization of hadronic
and boosted $W$ boson using jet substructure technique at
14~TeV LHC. Two different methods, {\it viz.} N-subjettiness
and Soft Drop, were used to find the subjets, which are
approximately considered to be the two hadronic decay
products of $W$, inside boosted $W$ jets. These subjets were
then used to find the distribution of $p_\theta$ and $z_j$
to prepare the templates of longitudinally and transversely
polarized $W$. We then used these templates to find the
fractions of different $W$ polarization in a mixed sample to
a relatively good accuracy.}

\maketitle
\vspace*{-1mm}
%%%%%%%%%%%%%%%%%%%%%%%%%  Main Body  %%%%%%%%%%%%%%%%%%%%%%
\section{Introduction}\label{sec:introduction}
The study of fundamental interactions between elementary
particles is the primary goal of particle physics. The probe
to these interactions is usually done via different types of
scattering processes. Although phenomena naturally occurring
in our surroundings involve scattering and reveal a very
high amount of information, dedicated man-made experiments
give better opportunity to probe such interactions in a
controlled way. Today's advanced colliders are the types of
experiments which reveal great information about the
fundamental interactions of nature. These high energy and
highly luminous colliders provide us ample opportunities to
study the fractions of longitudinal and transverse
polarization of heavy particles emerging either from the 
Standard Model (SM) or from beyond the SM (BSM) scenario
(like supersymmetry~\cite{MARTIN_1998} or composite Higgs
models~\cite{csaki2018tasi} where in some cases the heavy
resonance decays to a pair of essentially longitudinally
polarized $W$ or $Z$ bosons\,\cite{Kilian_2015,Kilian_2016}).
In the SM, it is the study of polarization fractions the
heavy bosons are particularly compelling since it reveals
the true nature of the electroweak symmetry breaking. The
$WW$ production via vector boson fusion (VBF)
production~\cite{Han_2010,Brehmer:2014dnr} tends to give
longitudinal $W$ bosons at the high energies. This is
because of the domination of the Goldstone nature of $W$ at
high energies. On the other hand, finding the fraction of
longitudinally polarized $W$ in a process discloses the
contribution of new physics in the process. The polarization
study of $W$ boson is therefore an important check for SM or
BSM scenarios. 

The most simplest way of examining the polarization of $W$
is via its decay products. Since the leptonic channels are
the cleanest channel at a collider, most of the
phenomenological studies of $W$ polarization has been
carried out in the leptonic channel.
Experimental collaborations at the LHC have also done the
same analysis and measured the polarization fraction of SM
$W$ boson. CMS collaboration has measured the value in
leptonic $W$+jet events\,\cite{Chatrchyan_2011} and ATLAS
collaboration has done it via semi-leptonic $t\bar{t}$
events\,\cite{2012}. Despite the clean channel in the
leptonic decay modes, it has missing energy in it and hence
make the study little bit difficult. On the other hand, in
the hadronic decay modes of $W$ both jets can be observed
and the study polarization does not have the ambiguity of
the missing energy. However, signals in the hadronic modes
are always tricky to separate them from the huge QCD
background in a collider, especially in a hadron collider.
In addition, other effects like pile-up (PU), underlying
event (UE) add another level of difficulty to the study via
the hadronic modes. However, better understanding of such
effects and recent advancements in mitigating these effects
allow us to study polarization in hadronic channel as well.
Although hadronic channel still is not at per with the
leptonic channels, it may complement the leptonic modes and
help us in gathering little more information from the
collider. This work is an attempt to improve on the existing
proposals on the study of polarization of $W$ via hadronic
channels, although there are still scope of improvements in
this direction. 

The interesting developments in this direction makes use of
machine learning or jet substructure based analysis. In the
jet substructure technique, a new variable $p_\theta$ has
been proposed in Ref.\,\cite{de2020measuring}. In this
reference, the authors showed that this variable is a proxy
for the variable $\cos\theta$, where $\theta$ is the angle
between the propagation direction of $W$ and one of the
decay product in the rest frame of $W$. The variable
$p_\theta$ can be reconstructed from the energy of the two
subjets inside the $W$ fatjet. The reconstruction of the
variable crucially depends on how accurately the two subjets
have been identified. In Ref.\,\cite{de2020measuring},
N-subjettiness was used to find the two axis of the subjets
after the grooming via Mass Drop
tagger\,\cite{Butterworth:2008iy}. However, in this work, we
showed that the polarization study using
N-subjettiness~\cite{Thaler_2011} can be improved if we do
not use any grooming method especially in the region of
$p_\theta\to 1$. We also used Soft
Drop\,\cite{Larkoski:2014wba} tagger to find the subjets
which also yields a quite decent results.

This article is organized as follows. We briefly discuss
about the polarization states of $W$ boson in
Section~\ref{sec:polarization}. Jet substructure and study
of polarization using jet substructure are discussed in
Section~\ref{sec:JSS-pol}. The template models and
calculation of variables are described in
Section~\ref{sec:template}. Section~\ref{sec:result}
discusses the main result of our study and finally we
summarize our work in Section~\ref{sec:summary}. 

%%%%%%%%%%%%%%%%%%%%%%%%%%%%%%%%%%%%%%%%%%%%%%%%%%%%%%%%%%%%
\section{$W$ Boson Polarization} \label{sec:polarization}
A massive particle with spin $j$ has a total of $2j+1$
polarization (helicity) states. However, distinguishing
among these polarization states is a difficult task in
itself. On the other hand, the study of polarization states
tells us about the interaction a particle has gone through.
For example, polarization study can reveal whether an
interaction is parity conserving or violating or the
underlying structure of the interaction the particle has
gone through. The same can be true for charge conjugation
or time reversal symmetry. In this work, we will be focusing
on the polarization states of a spin one particle, namely
$W$ boson. This spin one particle has 3 polarization states
and they are one longitudinal and two transverse
polarization states. Longitudinal and transverse
polarization states are identified with the eigenstate of
$\hat p\cdot\vec J$, where $\hat p$ and $\vec J$ are
3-momenta unit vector and angular momenta vector
respectively. Longitudinal polarization states are those
which has eigenvalue 0, while the transverse states are
those which has eigenvalue $\pm1$. The angular distribution
of the decay products of the $W$ boson will depend on the
polarization state it is in.

\begin{figure}
\includegraphics[width=0.99\textwidth]{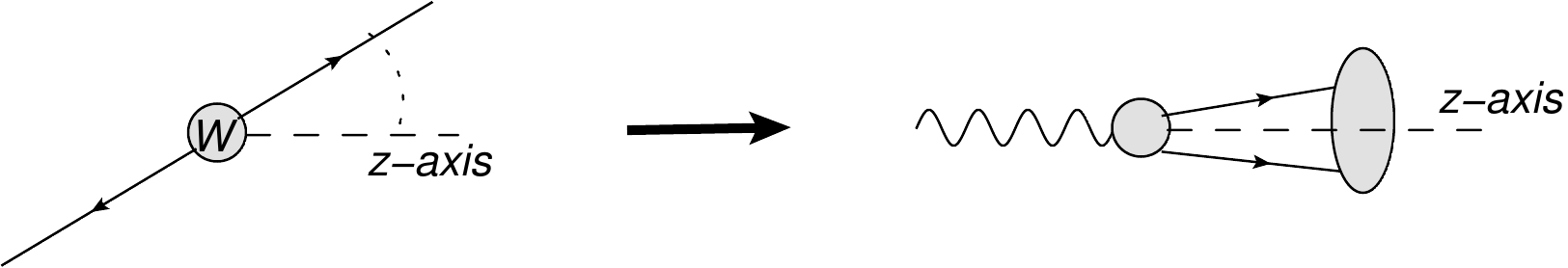}
\caption{Sketch of two body decay of $W$ in (left) its rest
frame and (right) in the lab frame.}
\vspace{-4cm}
\hspace{2.8cm}$q$

\vspace{-0.2cm}
\hspace{3.9cm} $\theta$

\vspace{-0.5cm}
\hspace{5.9cm} boosted
\hspace{2.0cm}$W$

\vspace{-0.8cm}
\hspace{12cm} $q$

\vspace{0.8cm}
\hspace{12cm}$q'$

\vspace{-0.4cm}
\hspace{1.2cm}$q'$

\vspace{1.5cm}
\label{fig:Wdecay}
\end{figure}

One of the most popular way to study polarization of a
particle, that can decay, is via the angular distributions
of its decay products. For the case of massive $W$ boson
which has two different types of polarization states, {\it
viz.} longitudinal and transverse, polarization of decaying
$W$ boson can be determined using its two-body decay
products. If a $W$ decays to two massless particles $q$ and
$q'$ in the lab frame, then one can boost back to its rest
frame with the $z$-axis to be taken along the propagation
direction of $W$ boson in the lab frame. In the rest frame,
one then can measure the angle between the $z$-axis and one
of the decay product as $\theta$. This has been depicted in
Figure~\ref{fig:Wdecay}. The decay of $W$ in its rest frame
is depicted in the left panel and the same in the lab frame
is depicted in right panel of the figure. When integrated
over the azimuthal angle in the rest frame of $W$, the
angular distribution of one of the decay product in the rest
frame of $W$ can be expressed as
\begin{align}
\frac{1}{\sigma}\frac{d\sigma}{d\cos\theta} &= f_0\frac{3}{4}\sin^2\theta + f_- \frac{3}{8}\left(1+\cos\theta\right)^2 + f_+\frac{3}{8}\left(1-\cos\theta\right)^2\\
&= f_0\frac{3}{4} (1-\cos^2\theta) + f_T \frac{3}{8}\left(1+\cos^2\theta\right) + f_D\frac{3}{4}\cos\theta  
\end{align}
where $f_{0,\pm}$ are fractions of different polarization
states present in $W$ sample and $f_T=f_++f_-$ is total
transverse polarization fraction and $f_D=f_--f_+$. We
should note that, in practical cases, all the helicity
states of the $W$ will interfere with each other~\cite{Buckley_2008,Buckley_2008_2,Ballestrero_2018}
to give rise to the final distribution. Integration over the
full decay azimuthal angles for $W$ boson decay eliminates
the interference terms although some applications of the
cuts (like maximum $\eta$ cut on hadrons or jets) will
reinstate some of the interference terms between the
different polarization states of the $W$
boson~\cite{Mirkes_1994,Stirling_2012,Belyaev_2013}.

By limiting ourselves to a measurement of $|\cos\theta|$
and $f_-=f_+$, the anticipated distribution is given by
\begin{align}
\frac{1}{\sigma}\frac{d\sigma}{d|\cos\theta|} &= f_0\frac{3}{2} (1-|\cos\theta|^2) + f_T \frac{3}{4}\left(1+|\cos\theta|^2\right) 
\label{final_eqn}
\end{align}

The variable $|\cos\theta|$ is defined in the rest frame of
$W$ while the $W$ is produced in lab frame, which, in
general, is not the rest frame of $W$. So, a variable that
mimics the variable $|\cos\theta|$ but calculated in the lab
frame will of course be useful. In
Ref.\,\cite{de2020measuring}, one such variable has been
suggested
\begin{eqnarray}
|\cos\theta_*| = \frac{|\Delta E|}{|\vec p_W|}
\label{costheta}
\end{eqnarray}
where $\Delta E$ is the difference in energy of the two
decay products and $\vec p_W$ is the 3-momenta of $W$ in the
lab frame. In addition, we have also used another variable
$z_{j_*}$ (momentum balance of decay product) for our
analysis in the polarization study of $W$. The variable is
defined as
\begin{eqnarray}
z_{j_*} = \frac{p_T^\text{leading}}{p^W_T}
\label{zj}
\end{eqnarray}
is basically the ratio of transverse momenta $p_T$ of the
leading jet and the $W$ boson\,\cite{roloff2021sensitivity}.
In this case also, both the $p_T$ of the decay product and
the $W$ are measured in the lab frame. As we have done the
study in the boosted $W$ jet, the variable calculation in
the lab frame is much more useful in terms of its subjets
inside the fatjet.
The variables $\cos\theta_*$ and $z_{j_*}$, which are
reconstructed  from the subjets of boosted jets will be
represented as $p_\theta$ and $z_{j}$ respectively.
%===========================================================
%===========================================================

%%%%%%%%%%%%%%%%%%%%%%%%%%%%%%%%%%%%%%%%%%%%%%%%%%%%%%%%%%%%
\section{Jet Substructure based Polarization Study} \label{sec:JSS-pol}
As we already discussed that the theoretical distributions
of $\cos\theta$ for longitudinally polarized $W$ and
transversely polarized $W$ are proportional to
$(1-|\cos\theta|^2)$ and $(1+|\cos\theta|^2)$ respectively. Most
of the existing polarization studies of $W$ or $Z$ bosons
are done with leptonic final states. However, in the case of
$W$ boson, leptonic final state contains a neutrino. The
weakly interacting neutrino then does not leave any trace at
the detector. This makes the polarization study in leptonic
decay modes little bit difficult. On the other hand, the
hadronic decay modes of $W$ produce two jets at the
detector. This, in principle, should make the polarization
study of $W$ easier.
However, effective reconstruction of jets and then $W$s
makes it more difficult to study polarization of $W$ at a
collider. Moreover, elimination of the huge QCD background
at a collider, especially at a hadron collider, adds up to
another level of difficulty. However, recent developments
in the study of jets and their substructures eases some of
the jobs of finding jets or subjets inside a fatjet.
Although we will not report signal-background type of
analysis in this article, we will try to show that the
$\cos\theta_*$ distribution can be reproduced with a good
accuracy for longitudinal and transversely polarized $W$
boson using its hadronic decay channel when $W$ is boosted
and gives rise to fatjet.

If the momentum of the decaying $W$ boson is high enough,
its decay products tend to be collinear. In case of
hadronic decay modes, these decay products again will shower
and will form collimated objects including all the collinear
decay products. Now, a jet clustering algorithm may not be
able to distinguish between
these highly collinear decay products and will cluster these
collimated final states into a single jet. These jets are
popularly known as fatjets or boosted jets. These boosted
jets are of high interest in the study of boosted
topologies. In our study, we too considered boosted $W$
jets. Once the $W$ fatjet is found, the job remains is to
find subjets inside the fatjet effectively. In this study,
we have used two different methods to find the two subjets
inside the boosted $W$ jets. These two methods are described
in the next two subsections. 

%===========================================================
\subsection{Subjet using N-subjettiness}
The polarization study as described the $\cos\theta_*$
variable relies on the effectiveness of the construction of
the subjets inside the boosted jet W. There are quite a few
method by which the subjets inside a boosted jet can be
found effectively. One of which is
N-subjettiness\,\cite{Thaler:2010tr}. This construction
mainly relies on finding the axes of a given number of
subjets. This method essentially partition the whole jet
area into $n$ number of subjet area. The proper definition
is as follows. Let $J$ be a boosted jet and $a_1, a_2,
\cdots a_n$ are a set of $n$ axes inside the boosted jet,
then we define a quantity
\begin{eqnarray}
\tau_n = \sum_{i\in J}\min_{a_1, a_2,
	\cdots a_n}\left\{d_{i,1},d_{i,2},\cdots,d_{i,3}\right\}
\label{eqn:Nsub}
\end{eqnarray}
where $J$ represents the full jet and $d_{i,j}$ is the
distance between $i^\text{th}$ constituent of the jet and
$j^\text{th}$ axis. The minimization in Eq.~(\ref{eqn:Nsub})
is done over the choices of the direction of the axes.

The distance measure and axes choices are also not unique.
There can be several different choices of the axes as well
as the distance measure depending on the types of
information one wants to extract. An exhaustive list of
such choices has been given in Ref.\,\cite{Stewart:2015waa}.
For our purposes, we tried various axes choices and measures
implemented in Fastjet Contrib\,\cite{Thaler:2010tr}. We
will list down the most effective choices in the result
section. 

%===========================================================
\subsection{Subjet using Soft Drop}
Soft Drop grooming/tagging method~\cite{Larkoski:2014wba}
was proposed to groom away the soft and wide angle
constituent, which comes predominantly  from PUs or UEs,
inside a jet. We, too, used it as a groomer to groom away
the contamination coming from PU and UE. However, we will
use this to find the subjets inside the boosted jet also. To
explain this, we first explain the Soft Drop algorithm below.

\begin{enumerate}
	\item Go back to the last stage of jet clustering. Let $j_1$ and
	$j_2$ be the two subjets giving rise to the final jet $J$.
	\item Check for the condition: $\dfrac{\min\{p_{T_{j_1}}\!,
	\ p_{T_{j_2}}\}}{p_{T_{j_1}}+p_{T_{j_2}}} > z_\text{cut}
	\left(\dfrac{\Delta R(j_1,j_2)}{R_0}\right)^\beta$
	\item If the condition in 2. is satisfied, declare $J$ as
	the final groomed jet. Otherwise, discard the softer subjet
	and promote the harder one to $J$ and restart from step 1. 
\end{enumerate}

In this way, we get the final groomed jet. One may consider
the two subjets to be the subjets of a two-pronged jet. This
is good approximation since, for a two-pronged jet, the two
prongs clusters first and then the two prongs combines to
give rise to the final jet. As described in the algorithm
itself, $z_\text{cut}$ and $\beta$ are parameters of Soft
Drop groomer and $R_0$ is the radius parameter of the
clustering algorithm that was chosen to cluster the jet
before applying Soft Drop grooming method. One important
observation is that $\beta\to\infty$ or $z_\text{cut}=0$
returns the original jet. Here, these two parameters are
chosen suitably so that we achieve our goal. In this work,
we consider this as one of the method to find subjets of
the boosted jets we will be considering later. Here again,
we checked with a number of different choices of these two
parameters. The best choices for our purposes will be given
in the result section.

As we already mentioned that the two variables $p_\theta$
and $z_j$ has already been suggested in the
literature\,\cite{de2020measuring,roloff2021sensitivity}.
Our main objective of this report is to improve on the
analysis.
As we already explained that $p_\theta$ correctly reproduces
$\cos\theta_*$, where $\theta_*$ is the angle between the
two decay products of $W$ in its rest frame. In the limit
where $\theta_*$ is $\pi/2$, both the decay products make
almost same angle with respect to the boost axis of the
decaying $W$ and after the boost, in the lab frame, they
share almost equal energy inside the fatjet. However, if
$\theta_*\to 0$ or $\pi$, one decay product is parallel to
the boost axis while the other is anti-parallel to the boost
axis. Hence, after the boost the parallel one becomes highly
energetic and the anti-parallel one becomes soft and wide
angle. Because of this, finding the subjet effectively is
difficult in the case of $|\cos\theta_*|\to 1$. In our
study, we mainly focused on this region so the
discrimination between longitudinal and transverse $W$
bosons can be improved. We applied the above two methods to
improve upon the earlier studies available in the
literature.

%===========================================================
\section{Generation of Templates}\label{sec:template}
As discussed earlier, we are interested in reconstructing
boosted $W$ jet and want to separate the two differently
polarized boosted $W$ boson using the variables $p_\theta$
and $z_j$. In this regard, we need to calibrate two kinds of
samples, one, which can contain a fully longitudinally
polarized boosted $W$ bosons and another with fully
transversely polarized boosted $W$ bosons. For this purpose,
we used two specific interaction Lagrangians which can be
implemented in FeynRules\,\cite{Alloul_2014} and already
used in Ref.\,\cite{de2020measuring}.
%-----------------------------------------------------------
\subsection{Template model}
Longitudinally polarized $W$ bosons can be generated via a
fictitious scalar particle $\phi_s$ which have a Higgs-like
couplings to W bosons and gluons,
\begin{equation}
\mathcal{L}_s = c_s^w \phi_s W^\mu W_\mu + c_s^g \phi_s G^{\mu \nu} G_{\mu \nu}
\label{slag}
\end{equation}
$c_s^w$ and $c_s^g$ are the coupling constants. We can see
that the interaction of the scalar with $W$ bosons is
through a non-gauge invariant, renormalizable term and it
picks out longitudinal $W$ bosons at high energies as per
Goldstone equivalence theorem. That is why there will be a
small admixture of transverse $W$ bosons if we produce $W$
via the s-channel process through $\phi_s$. However, the
fraction of transverse $W$ they will be suppressed by a
fraction $\sim\!\frac{m^4_W}{E^4} \simeq 10^{-3}-10^{-4}$
for W bosons with energies of order $500$~GeV to $800$~GeV.

On the other hand, transversely polarized $W$ bosons can be
produced by using non-renormalizable dimension-5 interaction
terms for a fictitious pseudo-scalar field $\phi_{ps}$. It
can couple to $W$s and gluons via the terms like
\begin{equation}
\mathcal{L}_{ps} = d_s^w \phi_{ps} W^{\mu \nu} \tilde{W}_{\mu \nu} + d_s^g \phi_{ps} G^{\mu \nu} \tilde{G}_{\mu \nu}
\label{pslag}
\end{equation}

For the sample of longitudinal $W$, we produced events in
{\tt MadGraph5}\,\cite{Alwall:2014hca} at a centre-of-mass
energy $\sqrt{s}$ = 14~TeV via the process $p p \rightarrow
\phi_{s}$ with $\phi_{s} \rightarrow W^+ W^-$ and for the
sample for purely transverse $W$, the same type of
$s$-channel process considered with $\phi_{s}$ replaced by
$\phi_{ps}$. Since we will be studying the polarization of
one of the boosted $W$ jet, one $W$ boson was allowed to
decay hadronically and the other was forced to decay
leptonically during the event generation using
{\tt MadGraph5}.

From Eq.~(\ref{pslag}), it can be shown that the amplitude
for $W$ boson production from the pseudo-scalar vertex have
the form $\mathcal{M} \propto \epsilon_{\mu \nu \rho \sigma}
p_1^\mu \epsilon_1^\nu p_2^\rho \epsilon_2^\sigma$, where
$\epsilon_{\mu \nu \rho \sigma}$ is the fully-antisymmetric
tensor and $p_i$, $\epsilon_i$ represent the four-momentum
and polarization vector for the $i^\text{th}$ $W$. This form
in the amplitude helps us to get a purely transverse
polarization vectors. As we are interested in boosted $W$
region produced via heavy resonance decay, the mass of the
scalar/pseudo-scalar are chosen $\sim$1~TeV.

\subsection{Generation of Sample Events}
\label{sample_events}

The following procedures have been conducted to generate the
sample events for our analysis.

\begin{enumerate}
\item For both (longitudinal and transverse) the cases, we
generated 8 lakh events in each of the cases, with an
intermediate scalar and pseudo-scalar using
{\tt MadGraph5}\,\cite{Alwall:2014hca}. At the parton level
event generation, we demand that the $W$ bosons have $p_T >
300.0$~GeV for both the cases. In order
to get relatively pure longitudinal sample, the events with
production of $W$ boson via $\phi_s$ has a cut on momentum
of $W$ boson which is $> 500.0$~GeV. At this parton level,
we can construct the variables by using relations,
\begin{eqnarray}
|\cos \theta_*| = \frac{|\Delta E|}{p_W}\\
z_{j_*} = \frac{p_T^{leading}}{p_T^W}
\end{eqnarray}
Here $p_T^{leading}$ is the $p_T$ of our leading hadron,
$p_T^W$ is the $p_T$ of $W$ boson and $|\Delta E|$ is the
absolute value of the energy difference of two hadrons
generated from $W$ decay.

\item We then used {\tt Pythia8}\,\cite{Sjostrand:2014zea,
Sjostrand:2006za} to shower and hadronize these parton level
events generated by {\tt MadGraph5}. For the analysis after
showering and hadronization, we used the parton level events
without any cut and with underlying events turned on.

\item We then cluster the final state hadrons with $|\eta| <$
4.0 using the Cambridge-Aachen algorithm in {\tt
FastJet}\,\cite{Cacciari:2011ma,Cacciari:2005hq} with a jet
radius $R_0$ = 1.0 and use two different methods to
reconstruct two prongs from the fatjet as described in the
previous section. As we can see that the pure longitudinal
and transverse samples have differences in the distributions
in the plane of the two variables, $p_\theta$ and $z_j$.
Since we are expecting two subjets in a $W$ fatjet, we have
taken $n=2$ for the calculation of N-subjettiness variable
$\tau_n$ as defined in Eq.~(\ref{eqn:Nsub}) for all the cases.
The different choice of axes and distance measures in
N-subjettiness technique and different $z_{cut}$ and
$\beta$ cut in Soft Drop technique can be useful depending
on which type of polarization we wants to study. This will
be explained in details in the next section where best case
scenarios will be studies with different kind of setups.
\begin{itemize}
\item In the case where we are interested in reconstructing
longitudinal $W$, for N-subjettiness, we used `OnePass
General $E_T$ General $K_T$ Axes' choice with jet radius
$R_0$ = 0.2 and $p$ = 0.6 (the limiting cases are $k_t$ and
Cambridge-Aachen axes choices for $p=1$ and $p=0$
respectively)\,\cite{Catani:1991hj,Catani:1993hr}.
With this, we used `Unnormalized Measure'
with $\beta$ = 1.0\,\cite{Stewart:2015waa}.
\item When we consider transverse $W$ as our best case, for
N-subjettiness we used same axis and measure choice with a
different $p$ value, $p$ = 0.05. 
\item For longitudinal $W$ best case two Soft Drop
parameters are chosen as $z_\text{cut}=0.26$ and $\beta=1.0$
while for transverse $W$ analysis they are set at
$z_\text{cut}=0.09$ and $\beta$ = 2.1.
\end{itemize}
For all the cases the events are selected if the
reconstructed $W$ (fatjet) has a $p_T >$ 300.0 GeV 

\item Detector level simulation are done by using
Delphes\,\cite{de_Favereau_2014}. We then added pile-up
events by considering minbias pile-up. After that the
samples are reconstructed with the similar way as described
in the last point with the same choices of N-subjettiness
and Soft Drop parameters.

\item For all the cases the final samples are chosen with
the following tagging cuts on the $W$ jets,\\
$\bullet$ Mass cut: 60 GeV $< M_{W/J} <$ 100 GeV, where
$M_J$ is the mass of the fatjet.

\end{enumerate}

We will be carrying out the analysis at three different
levels, {\it viz.} (a) {\it	parton level}, (b) {\it pythia
level}, and (c) {\it delphes level}. {\it Parton level}
means the variables were calculated at the parton level
final state i.e. after event generation in {\tt MadGraph5}
while {\it pythia level} means the variables were calculated
after showering and hadronization by {\tt Pythia8}. The
variables calculated after detector effects using Delphes
are represented by {\it delphes level}. At the parton level
analysis, we expect to get the similar distribution as
Eq.~(\ref{final_eqn}) by constructing the variable
$\cos\theta_*$ and $z_{j_*}$ at the lowest order
in QCD process using Eq.~(\ref{costheta}) and
Eq.~(\ref{zj}). In pythia level calculations, the quarks
from $W$ boson decay undergo through showering and
hadronization process to give rise to hadrons as final state
particles. These final state hadrons are then clustered as
jets. For detector level simulation pile up effects are also
added with those and the jets are formed after the detector
simulation. As we are interested in boosted $W$ region, the
jets coming from $W$ decay are highly collimated and most
often ends up by providing a fatjet. To classify the fatjets
correspond to $W$ bosons we can use the mass cut described
before. Along with that, to reconstruct the variables, we
need to find out the subjets from the fatjet by using jet
substructure techniques. As discussed earlier that we used
N-subjettiness and Soft Drop for this purpose. After finding
out the two prongs from the fatjet we reconstruct the two
variables $p_\theta$ and $z_j$ as
\begin{eqnarray}
p_\theta = \frac{|\Delta E^{reco}|}{p_W^{reco}}\\
z_{j} = \frac{p_T^{leading,reco}}{p_T^{W,reco}}
\end{eqnarray}
where the superscript `reco' has been added to represent the
reconstructed values for the momenta and the energies. So
${p_W^{reco}}$ represents the magnitude of the momentum of
reconstructed $W$ which is nothing but the magnitude of
momentum of the fatjet ($|p_J|$) at the hadron level or at
the detector level reconstruction of the $W$ momentum.
$|\Delta E^{reco}|$ can be calculated by taking the
difference in the energies of the two subjets reconstructed
at pythia and delphes level. Similarly, $p_T^{leading,reco}$
is the $p_T$ of leading subjet inside boosted $W$ jet and
$p_T^{W,reco}$ is the transverse momentum of reconstructed
$W$.

\section{Result}\label{sec:result}

\subsection{Best case scenarios}
\begin{figure}[!h]
\includegraphics[width=\textwidth]{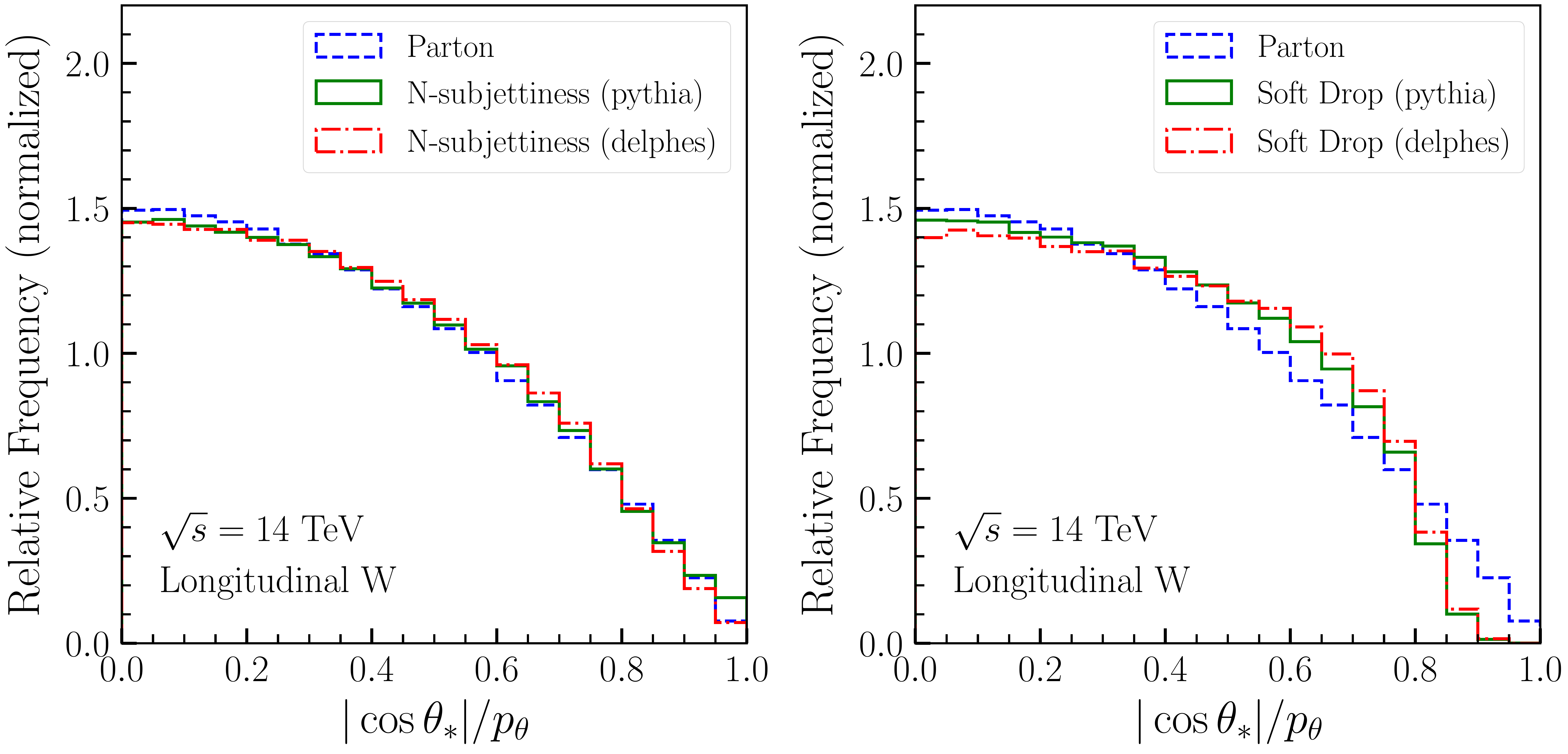}
\includegraphics[width=\textwidth]{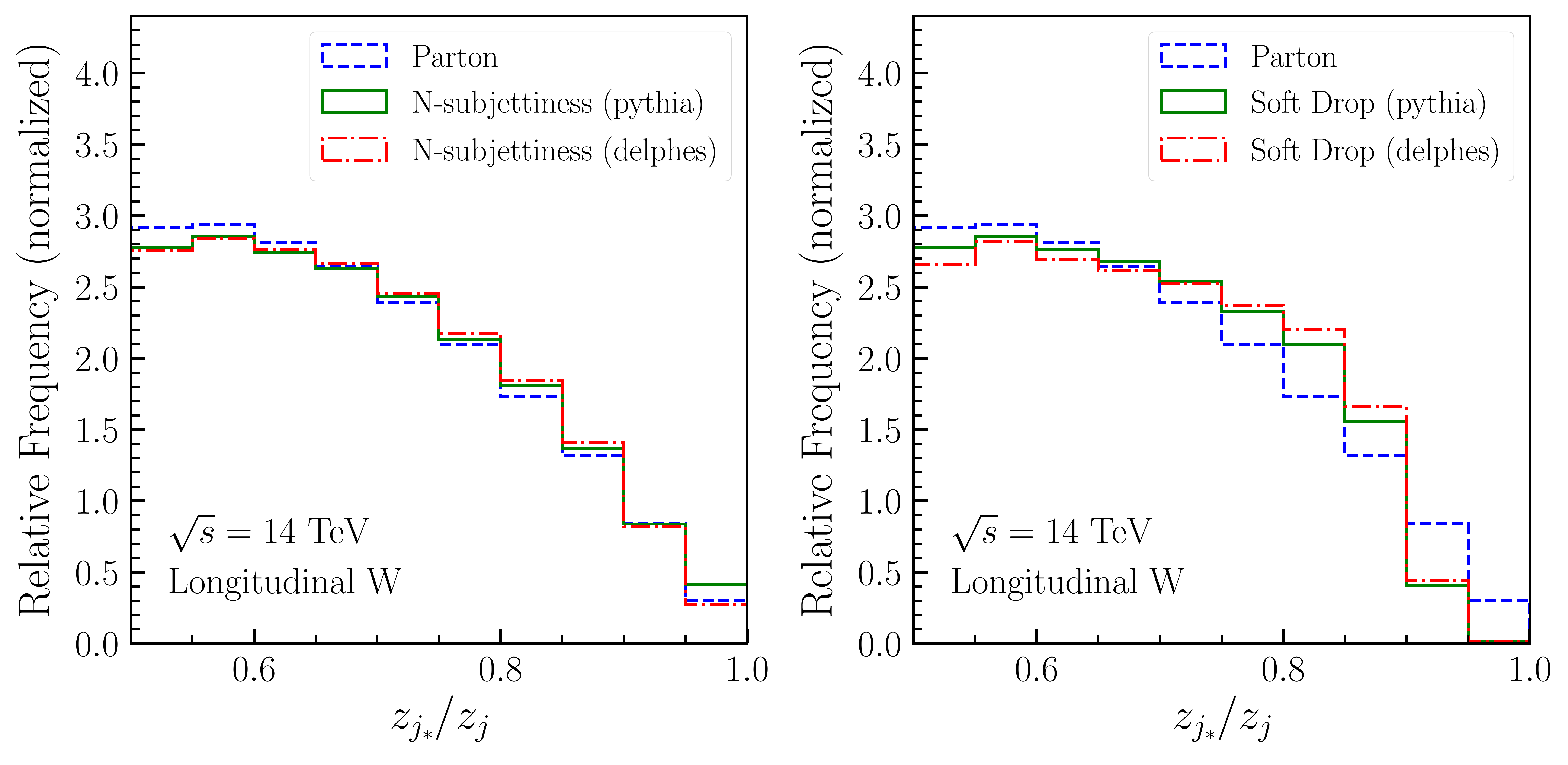}
\caption{Normalized distribution of angular variable
$p_\theta$ (upper panel) and momentum balance $z_j$ (lower
panel) for longitudinally polarized $W$. The distributions
are shown for parton level (blue dashed), pythia level
(green solid) and delphes level (red dash dotted) analysis
with N-subjettiness (left panel) and Soft drop (right panel)
techniques to find the subjets.}
\label{fig:match-longi}
\end{figure}

As we mentioned earlier that the study of polarization of
$W$ boson depends on how accurately the two subjets inside
these boosted jets can be reconstructed. In order to get
close enough distribution of $p_\theta$ to $|\cos\theta|$,
we have varied different parameters of Soft Drop and
N-subjettiness. We first did a thorough scan over these
parameters to get a good match to the theoretical
distribution of $|\cos\theta|$. We did not carry out the
matching for $z_j$ variable with $z_{j_*}$ since these two
variables are highly correlated. In our study, we found that
we need to take different values of the parameters for
longitudinal case than the transverse case.
We show these matching in Figure~\ref{fig:match-longi} for
longitudinally polarized $W$ (the Lagrangian is described by
Eq.~(\ref{slag})). As mentioned earlier that we carried out
the analysis at three different levels, {\it viz.} (a) {\it
parton level}, (b) {\it pythia level}, and (c) {\it delphes
level}. In all the panels of Figure~\ref{fig:match-longi},
blue dashed, green solid and red dash dotted histograms
represent the distributions of variables for parton level,
pythia level and delphes level analysis respectively. We can
see that both N-subjettiness analysis and Soft Drop analysis
provides good matching for the longitudinal case for both
the variables $p_\theta$ and $z_j$. The distribution is
little off near the value 1. The reason for this is that one
of the subjet is very soft in that region and hence it is
difficult to reconstruct that soft subjet effectively in
that region.

\begin{figure}[!h]
\includegraphics[width=\textwidth]{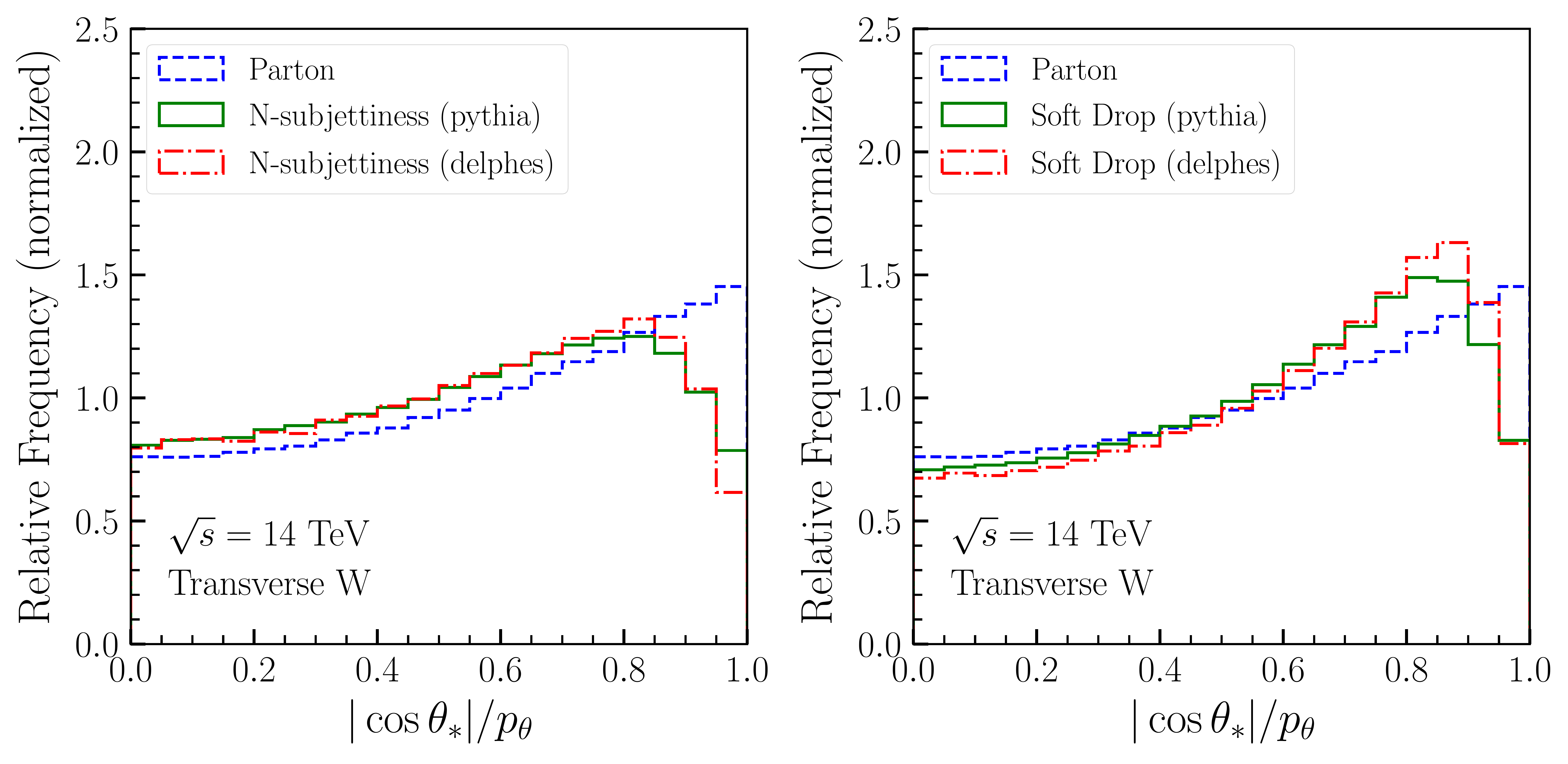}
\includegraphics[width=\textwidth]{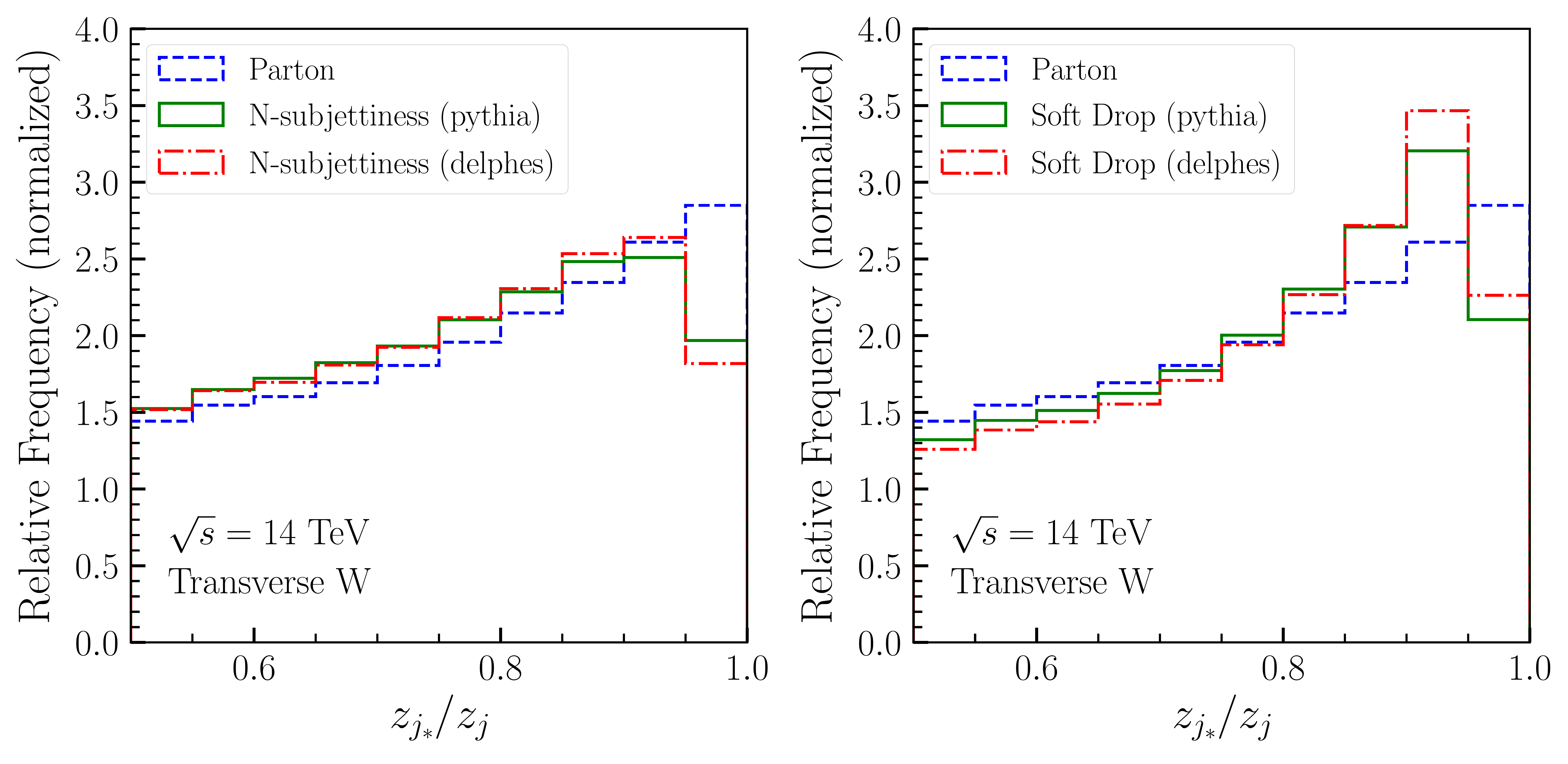}
\caption{Normalized distribution of angular variable
$p_\theta$ (upper panel) and momentum balance $z_j$ (lower
panel) for transversely polarized $W$. The convention for
the colours and the level are similar to
Figure~\ref{fig:match-longi}.}
\label{fig:match-trans}
\end{figure}

The same analysis has been done for the transverse case also. Figure~\ref{fig:match-trans} shows the distributions of the
same variables for the parton, pythia and delphes level
analyses. The conventions (colour, label etc.) are similar
to that of Figure~\ref{fig:match-longi}. We see the same
feature here again i.e. the distribution is not very
accurate near 1 because one of the subjet is very soft here. 

As we mentioned that the parameter choices for best case
scenarios are different for the longitudinal case than the
transverse one. We tabulate the parameter choices for best
case scenarios for both the longitudinal and transverse
cases in Table~\ref{tab:best}. The reason for the different
choices are quite clear from the distribution of $p_\theta$
as well as $z_j$ as shown in Figure~\ref{fig:match-longi}
and Figure~\ref{fig:match-trans}. The distributions peak
near 0 for the case of longitudinal $W$ whereas they peak
near 1 for the case of transverse $W$. For the case of Soft
Drop as a method of finding the subjets inside boosted $W$,
we need very soft $z_\text{cut}$ in order to keep the softer
subjet of the final jet. As in the case of transverse $W$,
we need to the peak near 1, the $z_\text{cut}$ value should
be smaller than that of longitudinal $W$. However, $\beta$
parameter of is mostly independent of which type of
polarization is being dealt with.

\begin{table}
	\begin{center}
%\begin{tabular}{ll}
\begin{tabular}{ |p{3.5cm}||p{3.5cm}|p{3cm}|c|c|  }
 \hline
 \multicolumn{5}{|c|}{N-subjettiness} \\
 \hline
 & Axes choice & Measure choice & {$\beta$-value} & $p$-value \\
 \hline
 Longitudinal\hspace{0.2cm}$W$ Best~case~scenario & `OnePass~General $E_T$\hspace{0.2cm}General\hspace{0.2cm}$k_T$ Axes' & Unnormalized Measure &1.0 &  0.6\\
 \hline
 Transverse\hspace{0.2cm}$W$ ~~Best case scenario& `OnePass General $E_T$\hspace{0.2cm}General\hspace{0.2cm}$k_T$ Axes' & Unnormalized\hspace{0.2cm}Measure & 1.0 & 0.05\\
 \hline
\end{tabular}\\
\begin{tabular}{ |p{3.5cm}||c|c| }
 \hline
 \multicolumn{3}{|c|}{Soft Drop} \\
 \hline
 & $\beta$-value & $z_\text{cut}$-value \\
 \hline
 Longitudinal\hspace{0.2cm}$W$ Best case scenario &  1.0 & 0.26\\
 \hline
 Transverse\hspace{0.2cm}$W$ ~~Best case scenario &  2.1 & 0.09\\
 \hline
\end{tabular}

\end{center}
%\end{tabular}
\caption{Parameter choices for the two techniques used to
find subjets inside $W$ fatjet. These values of parameters
are taken to optimize the templates for longitudinally and
transversely polarized $W$.}
\label{tab:best}
\end{table}

\subsection{Separability}
We then tried to check the separability between the
longitudinal and transverse $W$ bosons study. We did this
in terms of Receiver Operating Characteristic (ROC) curves.
When there is difference in the distribution of a variable
coming from two different types of sources, we may try to
get a score of their separability via ROC curves. These
curves are usually drawn to show how much a particular
distribution can be rejected at what acceptance level of the
other. This is usually done for signal and background
analysis where our main aim is to accept signal and reject
background effectively. However, ROC can also give us a
sense of separability of two distribution. Although
longitudinal and transverse distributions are not signal and
background analysis, we have drawn their ROC curves to show
their separability via this method. If two distributions are
identical, the area under the ROC curves are 0.5. In case
there are separation between the two different distribution,
the area under the curve varies from 0.5 to 1 with 1 being
the completely separable. Hence, closer the value of area
under the ROC curve to 1, better they can be separated.

\begin{figure}[!h]
\includegraphics[width=0.48\textwidth]{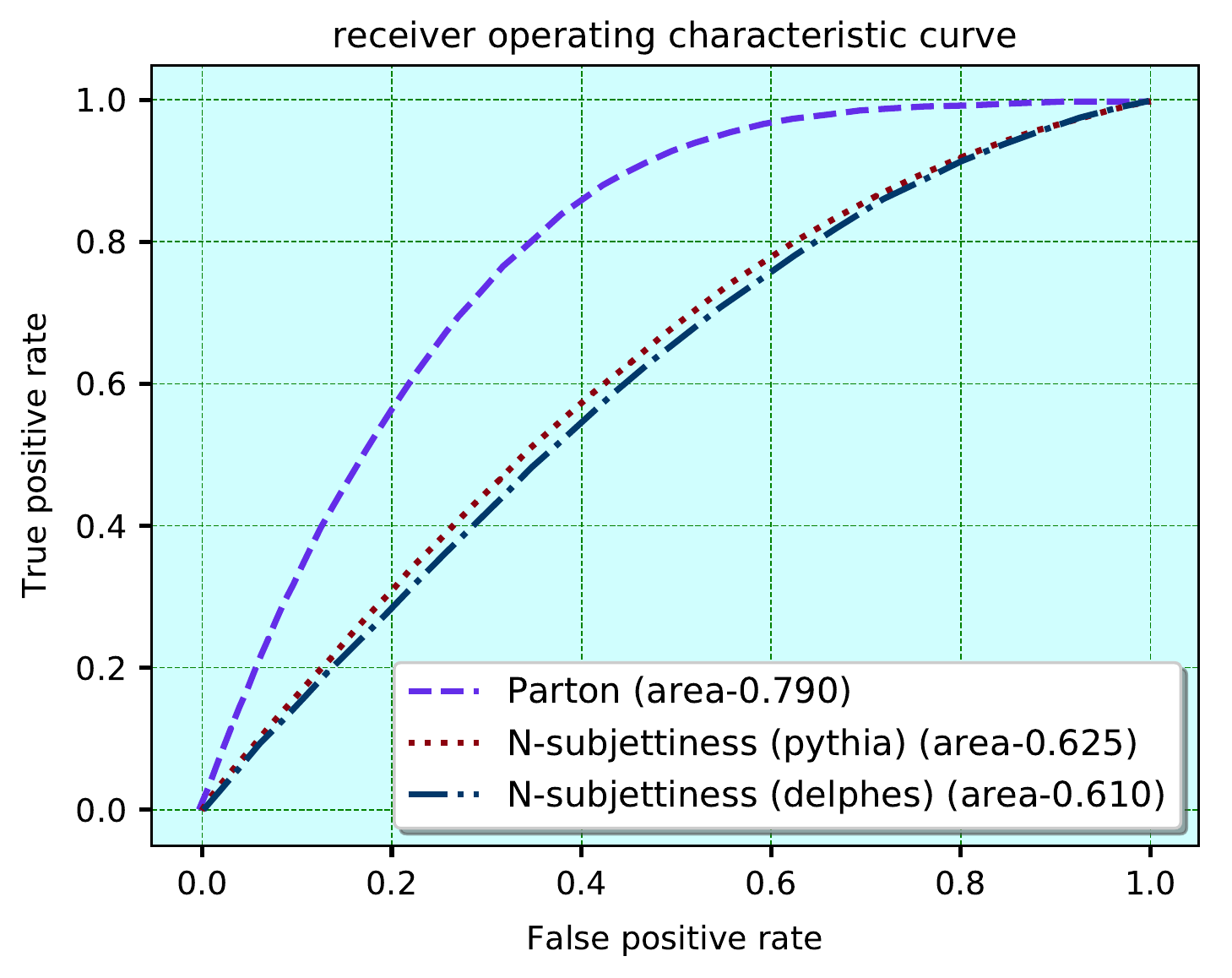}
\includegraphics[width=0.48\textwidth]{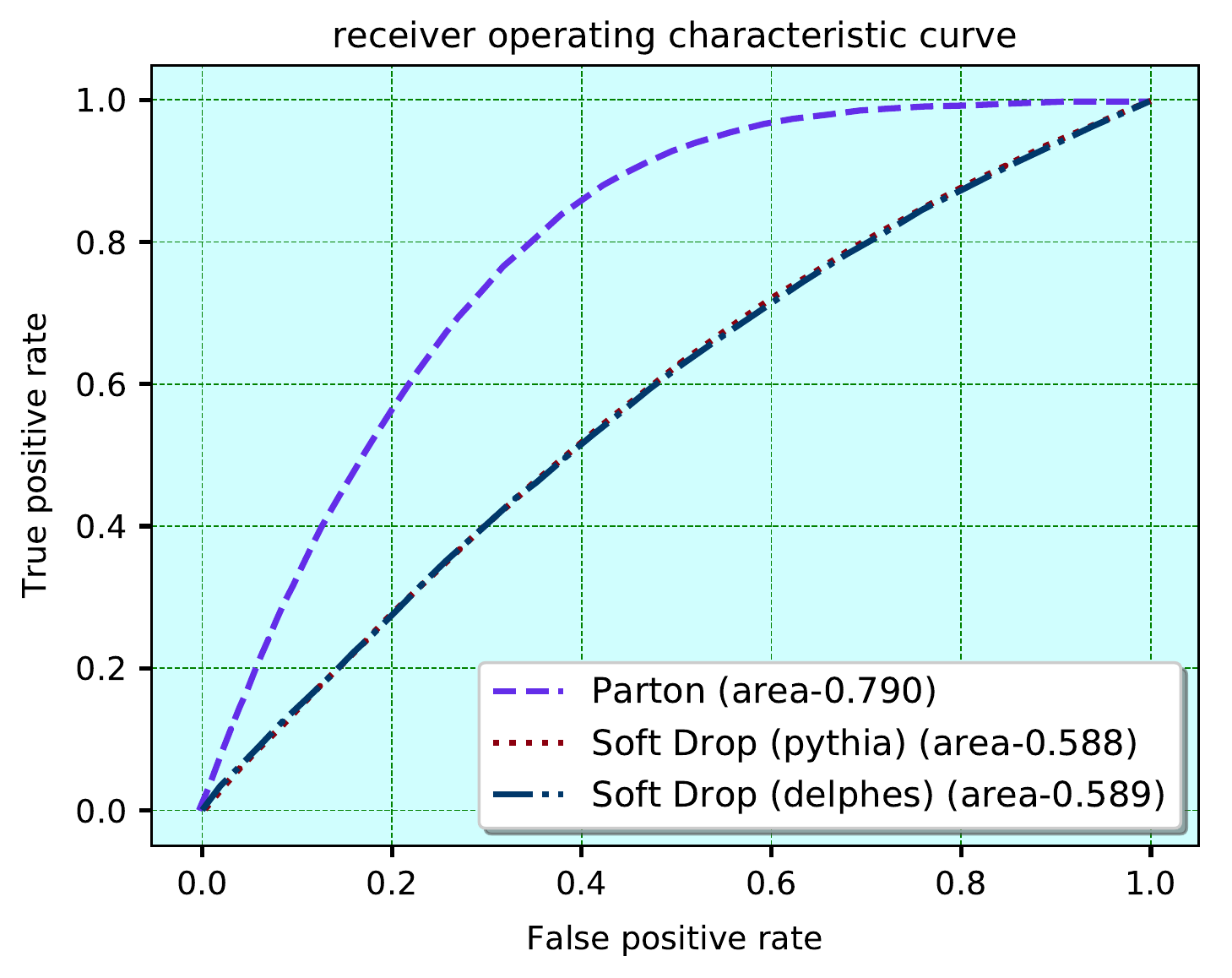}
\caption{ROC curves to illustrate the separability between
two templates, {\it viz}. longitudinal and transverse. The
parameter choices are taken to optimize the longitudinal
template.}
\label{fig:sep-longi}
\end{figure}

\begin{figure}[!h]
\includegraphics[width=0.48\textwidth]{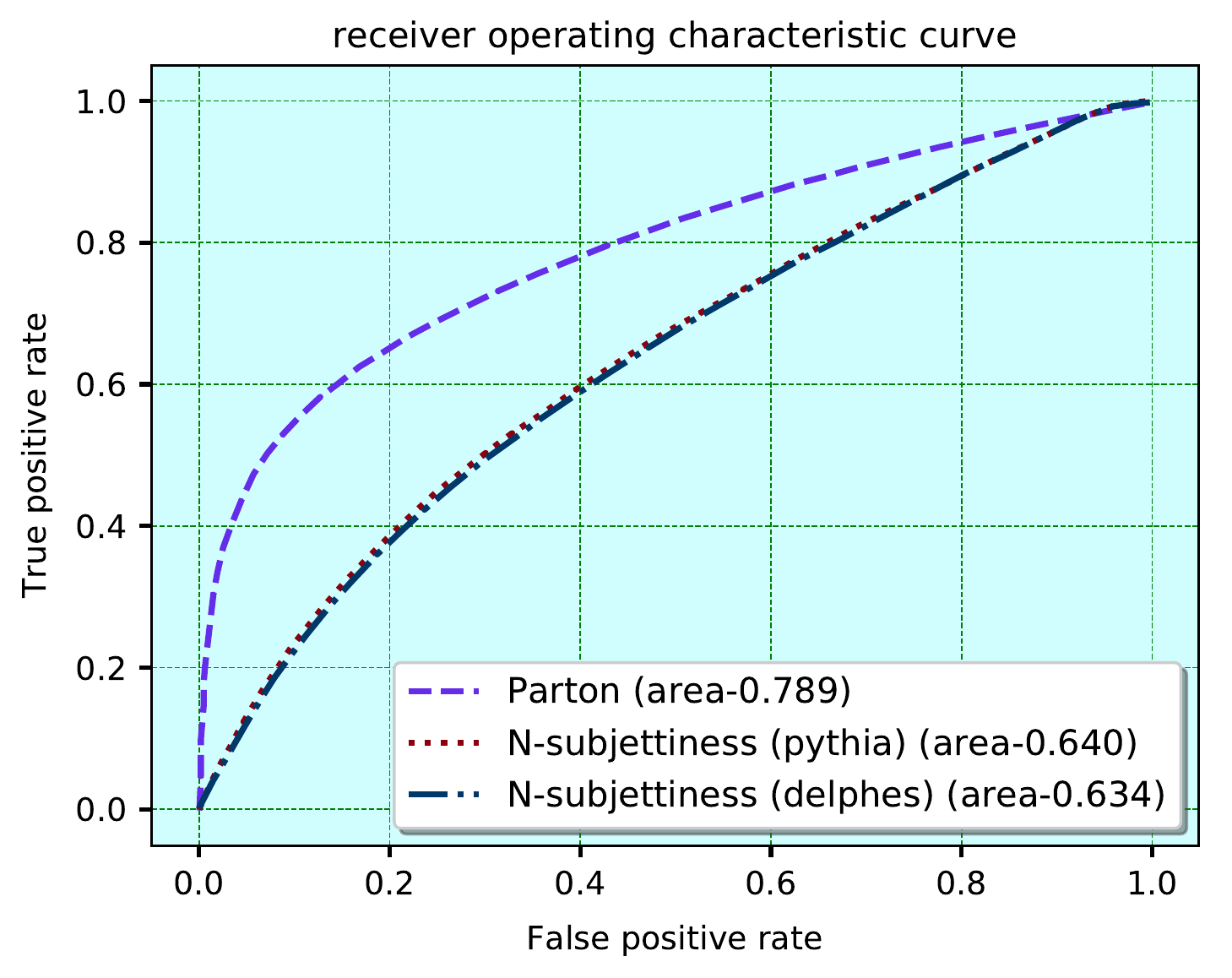}
\includegraphics[width=0.48\textwidth]{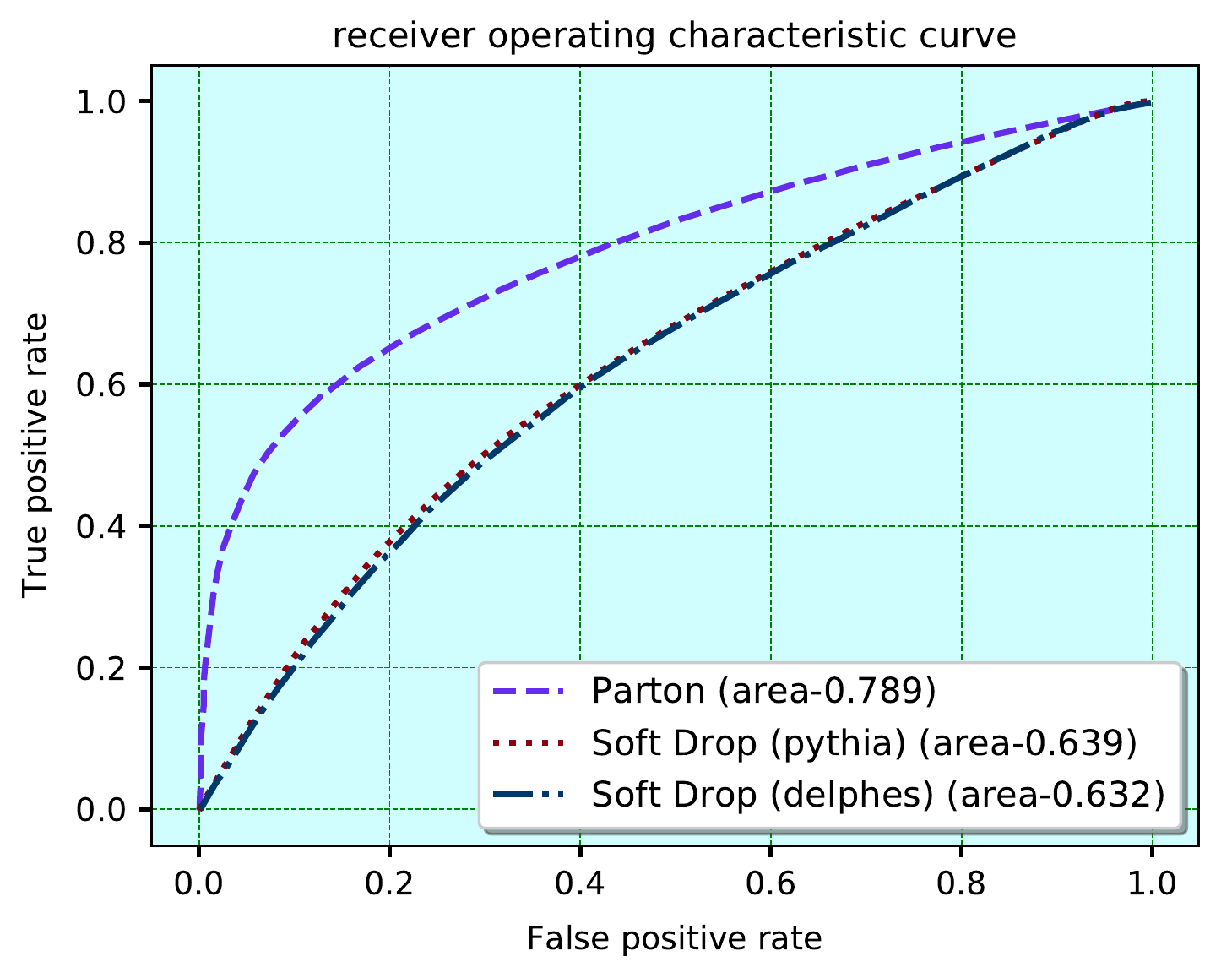}
\caption{ROC curves to illustrate the separability between
two templates, {\it viz}. longitudinal and transverse. The
parameter choices are taken to optimize the transverse
template.}
\label{fig:sep-trans}
\end{figure}

We consider both cases where, in one we want to get
longitudinally polarized $W$ event over the transverse one
(see Figure~\ref{fig:sep-longi}) and here we use the
parameter choice for the Longitudinal $W$ best case scenario
from Table~\ref{tab:best}. In another case we try to get the
transversely polarized $W$ dominated region over the
longitudinal one (see Figure~\ref{fig:sep-trans}) and here
we use the JSS parameters as per the transverse $W$ best
case scenario from Table~\ref{tab:best} using two feature
variables, $p_\theta$ and $z_j$. To get better separability,
we explore some recently developed techniques like Gradient
Boosted Decision Trees~\cite{Chen_2016}. The toolkit used
for Gradient boosting is XGBoost~\cite{Chen_2016}. For
gradient boosted Decision Tree method of separation, we
consider $\sim$5500 estimators and maximum depth of 4 where the
learning rate varies depending on the achievement to
separate longitudinal and transverse polarized samples at
different level of measurements (generator level, pythia
level or detector level) without overtraining. For parton
level analysis in both the cases our learning rate is 0.03
and we have used 80\% of our total dataset for training
purpose and 20\% for validation, where for other kind of
analysis our learning rate is 0.001 and we used 70\% of the
data to train our sample and 30\% for validation. 

In Figure~\ref{fig:sep-longi} we can see the ROC curve where
our signal is coming from longitudinally polarized $W$ decay
and our background is coming from transversely polarized $W$
decay. In left side plot we represent parton level analysis
and also pythia level and detector level analysis with
N-subjettiness technique. On the other hand in right side
plot similar things are represented by measuring with Soft
Drop technique. Alternatively Figure~\ref{fig:sep-trans}
shows the ROC curve where the signal is characterized by the
events with transversely polarized $W$ and the background
events are coming from longitudinally polarized $W$ decay.
Here also left and right side plot portrays the separability
using N-subjettiness and Soft Drop techniques respectively.
For all the cases, area under the ROC curve are listed in
Table~\ref{roc-table}.
\begin{table}
\centering
\begin{tabular}{ |p{3cm}||p{3.2cm}|p{3.3cm}|  }
 \hline
 \multicolumn{3}{|c|}{Measurement of separability (Area under the ROC curve)} \\
 \hline
 Analysis level/techniques& Longitudinal\hspace{0.2cm}$W$ Best case scenario & Transverse\hspace{0.2cm}$W$ Best case scenario\\
 \hline
 Parton   & \hspace{1cm} 0.790 & \hspace{1cm} 0.789\\
 \hline
 N-subjettiness (pythia)& \hspace{1cm} 0.625  & \hspace{1cm} 0.640\\
 \hline
 N-subjettiness (delphes) & \hspace{1cm} 0.610 & \hspace{1cm} 0.634\\
 \hline
 Soft\hspace{0.1cm}Drop (pythia) & \hspace{1cm} 0.588 & \hspace{1cm} 0.639\\
 \hline
 Soft\hspace{0.1cm}Drop (delphes) & \hspace{1cm} 0.589  & \hspace{1cm} 0.632\\
 \hline
\end{tabular}
\caption{Area under the ROC curves for different level of analysis}
\label{roc-table}
\end{table}

From Table~\ref{roc-table}, we can see that at parton level we can achieve better separability among all the cases. For Longitudinal best case scenario, at pythia level both N-subjettiness and Soft Drop achieve similar kind of separability where at delphes level, Soft Drop perform batter than N-subjettiness. On the other hand, for transverse best case scenario, all the techniques doing batter than previous case and at detector level analysis N-subjettiness doing batter than Soft Drop.

One of the possible shortcomings of these technique is over-training of the data sample where the training sample gives extremely good accuracy but the test sample fails to achieve that and we can see a noticeable difference in the ROC curve of training and testing case. We have explicitly checked that with our choice of parameters the algorithm we used does not overtrain.

\subsection{Template fitting}
We now use the above templates of longitudinal and transverse $W$ bosons to acquire information from a mixed sample. For this study, we first prepared sample events, which has admixture of longitudinal and transverse $W$ bosons in it. We then try to fit the this mixed sample events with the templates we generated earlier. Let $L(x)$ and $T(x)$ be the distribution for the variable $x$ for the two templates of longitudinally and transversely polarized $W$ bosons, respectively. These distributions are after the detector simulation and hence are not necessarily the same as the theoretical distribution. Let a mixed sample has the distribution $M(x)$ for the same variable $x$. The fraction, $\alpha$, of longitudinally polarized $W$ boson in the the mixed sample may be estimated by minimizing the following quantity. 
\begin{eqnarray}
\chi^2 = \sum_{i\,\in\,\text{bins}} \left|M(x_i)-\alpha L(x_i) - (1-\alpha) T(x_i)\right|^2
\end{eqnarray}
The minimization over the fraction $\alpha$ gives the estimate for $\alpha$ as
\begin{eqnarray}
\alpha = \frac{\sum_{i\,\in\,\text{bins}}\left(M(x_i)-T(x_i)\right)\left(L(x_i)-T(x_i)\right)}{\sum_{i\,\in\,\text{bins}}\left(L(x_i)-T(x_i)\right)^2}
\end{eqnarray}

In this part of the study, we used the Delphes level distributions as our templates for longitudinally and transversely polarized $W$. We then prepared mixed sample events with three different fraction of 25\%, 50\% and 75\%. We then tried to estimate the value of $\alpha$ for these mixed sample cases. The estimated values are presented in Table~\ref{tab:template-fit}. 

\begin{table}
\begin{tabular}{|c|c|c|c|}
\hline
~~Subjet found ~~ & ~~Template optimized~~ & ~~Sample prepared~~ & ~~Estimated $\alpha$~~ \\
 ~~using ~~& ~~best for~~ & ~~with $\alpha$~~ & \\
\hline
\hline
\multirow{6}{25mm}{\text{N-subjettiness}} & \multirow{3}{25mm}{Longitudinal} & 0.25 & 0.169\\
\cline{3-4}& & 0.50 & 0.454 \\
\cline{3-4}& & 0.75 & 0.731 \\
\cline{2-4}&\multirow{3}{25mm}{Transverse} & 0.25 & 0.239 \\
\cline{3-4}& & 0.50 & 0.499\\
\cline{3-4}& & 0.75 & 0.746\\
\hline
\multirow{6}{25mm}{\text{Soft Drop}} & \multirow{3}{25mm}{Longitudinal} & 0.25 & 0.182\\
\cline{3-4}& & 0.50 & 0.480 \\
\cline{3-4}& & 0.75 & 0.734 \\
\cline{2-4}&\multirow{3}{25mm}{Transverse} & 0.25 & 0.224 \\
\cline{3-4}& & 0.50 & 0.471\\
\cline{3-4}& & 0.75 & 0.703\\
\hline
\end{tabular}
\caption{Estimated fraction (4th column) and actual fraction (3rd column) of longitudinally polarized $W$ in a mixed sample in the two technique to find the subjets. }
\label{tab:template-fit}
\end{table}

We have done this analysis with both the subjet finding methods, {\it viz.} N-subjettiness as well as Soft Drop method, and for both the scenarios with the templates being optimized best for longitudinally and transversely polarized $W$. In this part of the analysis, we used only $p_\theta$ variable.  We can see from Table~\ref{tab:template-fit} that the fraction $\alpha$ can be estimated with relatively good accuracy for the case when template is best optimized for transversely polarized $W$ in the N-subjettiness subjet finding method.

\section{Summary and Outlook} \label{sec:summary}
To summarize, we have studied the polarization states of hadronically decaying boosted $W$ boson. We have considered 14~TeV centre-of-mass energy at the LHC in this study.  We first generated approximately pure longitudinal and transverse $W$ boson by taking appropriate template models and high enough $p_T$ cut to keep hadronic $W$ as a fatjet. The analysis was done using angular variable $p_\theta$ (a proxy for $|\cos\theta|$) and momentum balance $z_j$ calculated using momenta and energies of the two subjets inside boosted $W$s. We employed the technique of N-subjettiness and Soft Drop to find the two subjets inside $W$ fatjets. The analysis was done at three different levels {\it viz.} (a) parton level, (b) pythia level, and (c) detector level. The different parameters of N-subjettiness and Soft Drop were optimized to achieve better match to the parton level distribution of these two variables for longitudinally and transversely polarized $W$ bosons separately. Although the optimized values of the parameters are different in two differently polarized cases, the separability is quite good in these two cases. We then used the templates to get an estimate of the fraction of longitudinally polarized $W$ in a set of mixed sample events. The estimate are better for the case when the template is optimized for transversely polarized $W$ than the longitudinal case.

The primary improvement of this study is to find the subjets inside a fatjet with a relatively better accuracy. This techniques can be used in the studies where the subjets inside a boosted jet is needed to be found. Although we did not carry out signal-background analysis in this study, this technique can be used to do such type of studies. This improvement may be achieved other boosted objects like $Z$, $H$, $t$, or other heavy BSM particles.

\section*{Acknowledgements}
The authors would like to acknowledge support from the Department of Atomic Energy, Government of India, for the Regional Centre for Accelerator-based Particle Physics (RECAPP). TS would also like to acknowledge the useful discussions with Santosh Kumar Rai. 

\bigskip

%\bibliography{references.bib}
%\bibliographystyle{JHEP}
%\input{ref.bbl}
\providecommand{\href}[2]{#2}\begingroup\raggedright\endgroup
\end{document}